# SDG Target Interactions: The Philippine Analysis of Indivisible and Cancelling Targets

Vena Pearl Boñgolan*, Arian Allenson M. Valdez[1], Roselle Leah K. Rivera[2]

*bongolan@up.edu.ph
1) Department of Computer Science
University of the Philippines Diliman

2) College of Social Work and Community Development
University of the Philippines Diliman

**ABSTRACT**

The United Nations developed the 17 Sustainable Development Goals (SDGs), with 169 targets, to serve as a plan for solving the world's problems and achieving a more sustainable future. This is modeled as a graph with the targets as nodes, and with the interaction between targets as the edges of the graph. An exhaustive binary comparison is done to analyze the intra- and inter-goal target interactions, entailing over 14000 comparisons. The task is to assign a 'color' to an edge: positive (indivisible), zero (consistent) or negative (cancelling). This is done via a panel of experts who will evaluate the target interactions, through a web application that was developed for coloring the edges.

This is an on-going study, and so far, of the 1256 edges colored, only 36 are cancelling (negative), or 2.86%; more than 97% are positive interactions. So far, the "most negative" interactions involve: "Climate Change"; "Life Below Water"; "Peace, Justice and Strong Institutions"; and "Decent Work and Economic Growth".

Most useful for planning might be the "graph of beautiful targets" feature, which shows target with non-negative interactions, and how they connect to each other. These are the targets that may be worked on simultaneously, and currently has more than 130 nodes. This study can help researchers analyze which targets enable or constrain each other, what mitigation can be done to avoid conflicts, and can be configured for sub-national or regional study.

Web app at: **http://sdg-interactions.herokuapp.com/**

# Introduction

## 1.1. Background of the Study

In 2015, the countries which took part in the United Nations General Assembly came up with and agreed to the 2030 Development Agenda. This agenda states that all the parties involved (countries and stakeholders) pledge to act together to achieve a better future for the world. This includes sustainability in all aspects, ending world hunger and poverty, and protecting the planet from further degradation among many others. It was the 2030 Development Agenda that shed light on the 17 Sustainable Development Goals encompassing the 169 targets which were to be achieved by the year 2030. The SDGs aim to balance the economic, social, and environmental necessities present in the world. Regardless of a country's development status, the SDGs are to act as a blueprint to achieve prosperity and sustainable growth for future generations to come.

## 1.2. Statement of the Problem

The Philippines is known as a third world country with a declining development status. The county deals with problems like corruption, poverty, overpopulation, and etc. These problems hinder the improvement of the country's general status but as it was stated in the 2030 Development Agenda, the 17 Sustainable Development Goals were to serve as a universal guide for all countries regardless of their development status. The researchers aim to study the

Interaction of the 169 targets in the Philippine context, with the help of data gathered from professionals in different but relevant fields around the country. The data will be used to identify and analyze the reinforcing, neutral, and conflicting relationships of targets, be it inter or intra-goal. The mitigations of the targets that have conflicting relationships will be studied in order to find out the necessary actions which could be forwarded to different sectors to help in the effective and efficient way for the country to achieve the 2030 agenda.

1.3. Objectives

- To create an analysis of the SDG target interactions

- To identify the negative target interactions and their mitigations
- To rank the SDG targets based on number of negative interactions

1.4 Significance of the Study

An analysis of SDG target interactions can help guide the country in efficiently achieving these development goals through utilizing the potential of positive SDG target interactions and mitigating the problems of negative SDG target interactions. By focusing the analysis on the Philippine context, we can gain insight on how the implementations of projects for achieving the SDGs can be made more suitable for our country's situation.

1.5. Scope and Limitation

The current experts who were able to contribute to the study are mostly from (1) University of the Philippines College of Social Work and Community Development and (2) UP National College of Public Administration and Governance in the University of the Philippines Diliman. The data acquired up until June 2021 will be analyzed in this research. With the implementation of the system required for the study already completed, this paper will focus more on:

- improvements made on the existing implementation

- the results and discussion based on the current data.

**2. Methodology**

Many have attempted to evaluate SDG target interactions through experts' judgement (International Science Council, 2017). Using the same method for evaluating SDG target interactions, this study continues efforts to evaluate all the target interactions (Abaja, Concepcion and Bongolan, 2019; Fernandez and La Rosa, 2020). We model the 169 targets as nodes in a graph, while the target interactions are the edges connecting all the nodes: 169*168/2 or 14,196 of them! A web application was developed to facilitate this process of constructing this graph. Experts must choose the SDGs that are aligned with their respective fields. The system will generate a series of SDG target pairs from their selected SDGs, and the expert must score the interaction using the 7-point scale (International Science Council, 2017). The goal is to be able to score all 14,196 target interactions.

2.1. Front-End Development: React

React (Abramov, D.) , was used for the front-end development of the project. It is a JavaScript library for making user interfaces on mobile and web applications. It was used to manufacture and design user views, admin views, and functionalities that are essential to the project. Some functionalities involve submitting the survey, reviewing answers, choosing goals, etc. which depend whether the person is an admin or a user. Levels of authority were also strictly implemented as only admins would be allowed to view certain pages. Necessary privacy measures were also observed. The developers also made use of **Reactstrap** to integrate the necessary components from **Bootstrap**. **Axios** was also used to be able to connect to the database. It is a Promise-based HTTP client that is widely used on the web.

The rendering of the graph for the project is developed using the **react-d3-graph** library. It was used by the developers to be able to create and tweak the graphs used for the application.

The following table shows the user stories for the user and administrator, with new functionalities highlighted:

| *As an Admin I can* | *As a User I can* |
|---|---|
| Confirm pending users | Add my goals |
| View Dashboard | View Front Page |
| View Graph | Log In |
| View Menu Bar | View Menu Bar |

| |
|---|
| View All Answers Review Answers |
| View Users View Settings |
| Sign Up |
| View Survey |
| View Tabs |
| View Ugly Targets |
| View Beautiful Targets |
| Select Specialist Goal |

2.2. Back-end Development: Flask

The Back-end of the system serves up an API and collects, saves, and processes the data. The backend is developed on top of Flask. The API requests handle user creation and data gathering.

PostgreSQL is the database system used, interfaced with the Python SQL toolkit, SQLAlchemy. There are four models in the system: the User model, the SDG model, the User to SDG model, and the Survey Answers model. All necessary information of the users are stored in the User model. Besides their login credentials, it also contains information to check if they are qualified to participate in the study like educational attainment and years of experience. A user may only start answering the survey upon the approval of an admin. Their corresponding curator, stored in the database as their contact person, will be notified of their sign-up so they can be approved. The other information stored concerns additional information about the users like affiliations and their preferences like whether they would like to be acknowledged in the research. For the reference of the users, the target descriptions are stored in the SDG model. The User to SDG model stores the users' chosen SDGs based on their expertise. From chosen SDGs, the SDG target interaction pairs are generated and linked to their user, and these are stored in the Survey Answers model. The APIs created handle the processes to be discussed in Data Gathering.

2.3. Data Gathering

The web application developed is the avenue for data gathering. To ensure the credibility of the SDG target interaction scores, the experts were required to have at least 5 years of experience in their field. Each participating expert must create an account on the web application to start scoring the SDG target interactions. Most of the experts are brought in by the study's curators, who will then have to approve their accounts before they can access login, further ensuring their credibility. Figure 1 shows the sign-up interface where all this information is required. Upon logging-in, the users were asked to choose the SDGs of their expertise. The system will generate SDG target pairs from the users' chosen SDGs.

Each SDG target pair represents the interaction between the two SDG targets, and the users will have to score each pair positive or negative using the 7-point scale. They were also given the opportunity to give an explanation for their score, however if a negative score was given, the explanation is required as well as mitigations to avoid the negative interaction. Users have the option to skip certain SDG target interactions that they do not wish to answer at the moment, but they will always be able to try to answer them.The SDG target pairs are bound to their user, and they will not be reassigned to a different user. Each SDG target pair can only be scored once, which means the score given by the users is final. The scores will be saved in the database upon submission of the answers.

The current progress of the data gathering can be viewed even without an account on the website. Users can select at most two SDGs and a graph will be generated showing the target interactions as edges of each SDG target as nodes. A sample of the network of SDG target interactions can be seen on Figure 4. The nodes are color-coded based on their SDG color. They are also labeled with their SDG and target number. The positive target interactions are the blue edges, the negative are the red edges, and the neutral ones are black. The gray edges are the target interactions yet to be colored by the experts.

2.4. Ugly and Beautiful Targets

We also report obtaining the negative target interactions as 'ugly' targets, and exclusively positive target interactions as 'beautiful' targets, and make these reports viewable on the website.

2.5 Longest Path of "Beautiful Targets"

We are trying to render the longest path of "Beautiful Targets" on the website, which is the longest path that only consist of positive target interactions, and avoids nodes or targets that conflict with other targets; this can be obtained by using a streaming longest path algorithm (Kleimann, L., Shielke, C. and Srivastav, A., 2016). We convert our undirected graph into a directed acyclic graph (already available as a graph), and we are currently working on the topological sort, which will allow us to render and obtain the longest path in a reasonable amount of time (linear).

## 3.Results and Discussion

The network of SDG target interactions currently has 1256 colored edges. Out of the 1256 colored edges, 983 are positive, 36 are negative, and 237 are neutral. Considering that only 1256 edges out of the 14196 total edges have been colored, there is still a need to get more responses from experts to obtain more credible data, however the current data can already give us initial information about the SDGs in the Philippine context.

One interaction we have looked at is 'Ugly Targets' and 'Beautiful Targets'. We have created a new page in our application which looks at these targets.

In the context of our study, Ugly Targets are targets that are conflicting or have a negative 'edge'.

| Score | | | |
|---|---|---|---|
| -3 | Show Info | 13.1 Climate Action | 14.C Life below Water |
| -3 | Show Info | 13.1 Climate Action | 14.5 Life below Water |
| -3 | Show Info | 16.B Peace, Justice, and Strong Institutions | 16.A Peace, Justice, and Strong Institutions |
| -3 | Show Info | 8.4 Decent Work and Economic Growth | 16.8 Peace, Justice, and Strong Institutions |
| -3 | Show Info | 8.2 Decent Work and Economic Growth | 13.3 Climate Action |
| -3 | Show Info | 8.1 Decent Work and Economic Growth | 11.6 Sustainable Cities and Communities |
| -2 | Show Info | 3.A Good Health and Well being | 16.1 Peace, Justice, and Strong Institutions |
| -2 | Show Info | 12.5 Responsible Consumption and Production | 17.15 Partnership for the Goals |
| -2 | Show Info | 5.3 Gender Equality | 10.6 Reduced Inequalities |
| -2 | Show Info | 8.2 Decent Work and Economic Growth | 12.3 Responsible Consumption and Production |
| -2 | Show Info | 1.5 No Poverty | 5.A Gender Equality |
| -2 | Show Info | 5.A Gender Equality | 16.A Peace, Justice, and Strong Institutions |
| -2 | Show Info | 3.1 Good Health and Well being | 3.6 Good Health and Well being |
| -2 | Show Info | 5.5 Gender Equality | 16.2 Peace, Justice, and Strong Institutions |
| -2 | Show Info | 5.1 Gender Equality | 13.1 Climate Action |
| -1 | Show Info | 5.6 Gender Equality | 5.2 Gender Equality |
| -1 | Show Info | 4.5 Quality Education | 4.7 Quality Education |

## Beautiful Targets

| Score | | | |
|---|---|---|---|
| 3 | Show Info | 9.4 Industry, Innovation and Infrastructure | 9.5 Industry, Innovation and Infrastructure |
| 3 | Show Info | 4.C Quality Education | 4.A Quality Education |
| 3 | Show Info | 8.9 Decent Work and Economic Growth | 9.3 Industry, Innovation and Infrastructure |
| 3 | Show Info | 2.A Zero Hunger | 5.A Gender Equality |
| 3 | Show Info | 1.2 No Poverty | 2.4 Zero Hunger |
| 3 | Show Info | 1.1 No Poverty | 10.4 Reduced Inequalities |
| 3 | Show Info | 12.4 Responsible Consumption and Production | 12.6 Responsible Consumption and Production |
| 3 | Show Info | 12.4 Responsible Consumption and Production | 12.5 Responsible Consumption and Production |
| 3 | Show Info | 2.5 Zero Hunger | 2.3 Zero Hunger |
| 3 | Show Info | 1.2 No Poverty | 1.5 No Poverty |
| 3 | Show Info | 1.3 No Poverty | 2.3 Zero Hunger |
| 3 | Show Info | 7.2 Affordable and Clean Energy | 16.A Peace, Justice, and Strong Institutions |
| 3 | Show Info | 1.B No Poverty | 2.2 Zero Hunger |
| 3 | Show Info | 1.A No Poverty | 1.4 No Poverty |
| 3 | Show Info | 1.4 No Poverty | 2.C Zero Hunger |
| 3 | Show Info | 1.A No Poverty | 2.B Zero Hunger |
| 3 | Show Info | 1.1 No Poverty | 2.B Zero Hunger |

Based on the ISC's 7-point scale, each SDG target interaction pair was given a score by the experts.

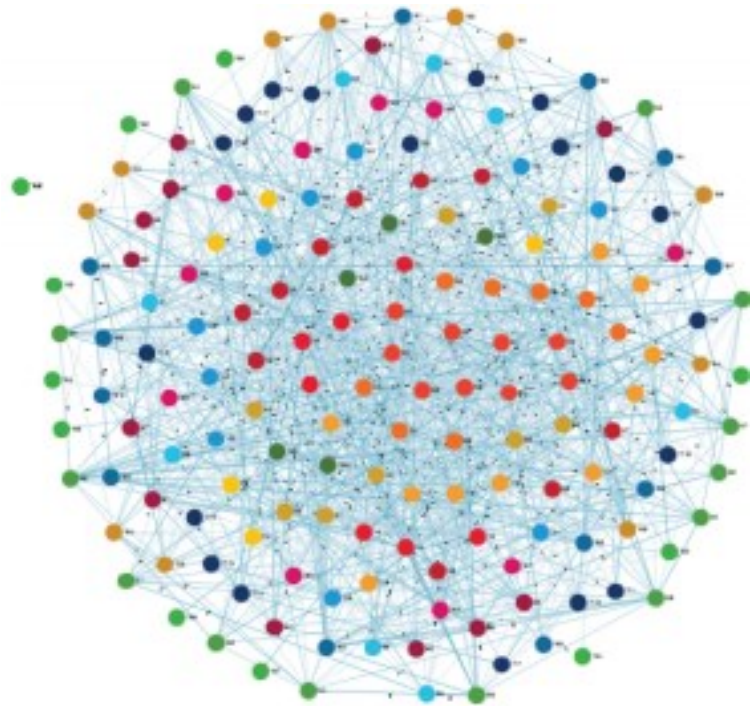

The Graph of Beautiful Targets is shown above.